\documentclass[pre,preprint,aps]{revtex4-1}
\usepackage{etoolbox} \newtoggle{memo} \newtoggle{showgit}
\toggletrue{memo}   
\togglefalse{memo}  
\newcommand{\note}[1]{\textcolor{red}{#1}}
\renewcommand{\note}[1]{} 
\togglefalse{showgit}  

\usepackage[utf8]{inputenc}
\usepackage{amsmath}
\usepackage{amsfonts}
\usepackage{amssymb}
\usepackage[breaklinks]{hyperref}
\usepackage{color}
\usepackage{xcolor}
\usepackage{siunitx}                
\hypersetup{
    colorlinks,
    linkcolor={red!50!black},
    citecolor={blue!50!black},
    urlcolor={blue!80!black}}
\usepackage{graphicx}
\usepackage{fancyhdr}               
\usepackage{listings}		        
\usepackage[version=3]{mhchem}      
\graphicspath{{fig/}}

\newcommand{\figurewidth}{3.4in}

\iftoggle{showgit}{
    \def\myfile{currentcommit.tmp}
    \openin0=\myfile\relax
    \read0 to \CurrentCommit%
    \closein0

    \newread\gitstatus
    \immediate\openin\gitstatus=uncommitted-files.tmp
    \read\gitstatus to \CurrentGitStatus
    \immediate\closein\gitstatus
    
    \newread\commitdate
    \immediate\openin\commitdate=commit-date.tmp
    \read\commitdate to \CommitDate
    \immediate\closein\commitdate
}{}

\fancypagestyle{body}{
    \fancyfoot{} 
    \fancyhead{} 
    \fancyfoot[C]{\thepage} 
    \setlength{\headheight}{14pt} 
}
\iftoggle{showgit}{
    \fancypagestyle{lastpage}{
        \fancyfoot{} 
        \fancyhead{} 
        \fancyhead[C]{\thepage}
        \fancyfoot[L]{commit \CurrentCommit}
        \fancyfoot[C]{\CommitDate}
        \fancyfoot[R]{\CurrentGitStatus uncommitted files}
        \setlength{\headheight}{14pt} 
    }
}{}

\begin{document}
\pagestyle{body}

\iftoggle{memo}{\preprint{NRL/MR/6790-16-9706}}{}
\title{Backward Raman Amplification in the Long-wavelength Infrared}
\author{L. A. Johnson}
\author{D. F. Gordon}
\author{J. P. Palastro}
\author{B. Hafizi}
\affiliation{U.S. Naval Research Laboratory, Washington, DC 20735} 
\date{\today}

\begin{abstract}
The wealth of work in backward Raman amplification in plasma has focused on the extreme intensity limit, however backward Raman amplification may also provide an effective and practical mechanism for generating intense, broad bandwidth, long-wavelength infrared radiation (LWIR).
An electromagnetic simulation coupled with a relativistic cold fluid plasma model is used to demonstrate the generation of picosecond pulses at a wavelength of \SI{10}{\micro\m} with terawatt powers through backward Raman amplification. 
The effects of collisional damping, Landau damping, pump depletion, and wave breaking are examined, as well as the resulting design considerations for a LWIR Raman amplifier. 
\end{abstract}

\maketitle

\iftoggle{memo}{
\tableofcontents
\newpage
}{}

\section{Introduction}

Chirped pulse amplification has provided access to intense, few cycle, near-infrared laser pulses for the last 30 years \cite{strickland1985compression}. 
It has been proposed that backward Raman amplification could provide similar access to the multipetawatt \cite{trines2011} or even exawatt regime \cite{malkin1999fast, malkin2000ultra}. 
While experiments have yet to reach the multipetawatt regime \cite{yampolsky2008demonstration,yampolsky2011limiting, turnbull2012simultaneous, turnbull2012possible,wu2016demonstration, ping2004amplification, cheng2005reaching, ren2008compact}, we propose that backward Raman amplification could be a practical source of long-wavelength infrared radiation (LWIR) with terawatt peak powers. 

The development of terawatt power pulses in the long-wavelength infrared is being driven by strong-field science, including advanced proton acceleration \cite{palmer2011monoenergetic}, high harmonic generation \cite{popmintchev2012bright}, mid-infrared supercontinuum generation \cite{pigeon2014generation,pigeon2015high}, and  nonlinear optics \cite{mitrofanov2015mid}.

There are several paths to high power LWIR: 
    optical parametric amplification (OPA), 
    difference-frequency generation (DFG),
    optical rectification (OR), 
    and lasing using \ce{CO2}.  
In principle, the Manley-Rowe relations limit the conversion efficiency to the ratio of the photon energies $\hbar \omega_\text{LWIR} / \hbar \omega_\text{NIR}$, about 10\%.
In practice, the conversion efficiency is an order of magnitude less, 
for example:
    optical parametric amplification becomes inefficient due to 
        absorption \cite{voronin2016modeling} 
        and group-velocity mismatch \cite{cerullo2003ultrafast};
    difference-frequency generation is limited by 
        phase-matching and optical nonlinearities \cite{sell2008phase};
    and optical rectification is limited because 
        it relies on the pump's spectral wings \cite{sell2008phase}.  
As a result, frequency down-conversion from the near-infrared to LWIR is inherently inefficient.

Existing high power \ce{CO2} amplifiers can create picosecond pulses with joules of energy \cite{polyanskiy2011picosecond,pigeon2015high} at wavelengths of \SI{9.4}{\micro\m} and \SI{10.6}{\micro\m}. 
However, these amplifiers require a combination of large high pressure systems, expensive oxygen isotopes, and a high power seed pulse to sufficiently broaden the \ce{CO2} gain spectrum for the generation of picosecond pulses. 
Furthermore, high gas pressure severely limits the system's repetition rate.   
If sufficient broadening is not achieved, these systems create pulse trains instead of individual pulses. 
Two attempts have been made to overcome the bandwidth limitations of \ce{CO2}, one used the negative group velocity dispersion in \ce{GaAs} with a beatwave to compress \ce{CO2} laser pulse trains \cite{pigeon2015high}, and the other by the combination of self-chirping and a conventional dispersive compressor \cite{pogorelsky2015bestia}.

This suggests that a compact \SI{10}{\micro\m} wavelength source capable of terawatt powers and picosecond durations would be a key scientific and technological development. 
Plasma-based, backward Raman amplification can provide the bandwidth that is difficult to achieve in a \ce{CO2} amplifier. 
This mechanism is self-phase-matched, and furthermore, optical nonlinearities in plasmas are weaker than in crystals. 
Additionally, the pumping \ce{CO2} laser can be narrow-band, which removes the need for operation at high pressure, and in turn, improves the prospects for high repetition rate operation.


In backward Raman amplification, a backward propagating seed pulse, with frequency and wavenumber $(\omega_1\approx\omega_0-\omega_p, k_1\approx-k_0)$, stimulates the coherent backscattering of a forward propagating pump pulse $(\omega_0, k_0)$ from a plasma (Langmuir) wave $(\omega_p, k_p\approx 2k_0)$. 
This enhances the plasma wave and further drives the energy from the pump to seed waves leading to exponential growth in the seed energy until pump depletion or another saturation mechanism occurs \cite{kruer1988physics}.

\begin{figure}[!ht]
    \centering
        \includegraphics[width=\figurewidth]{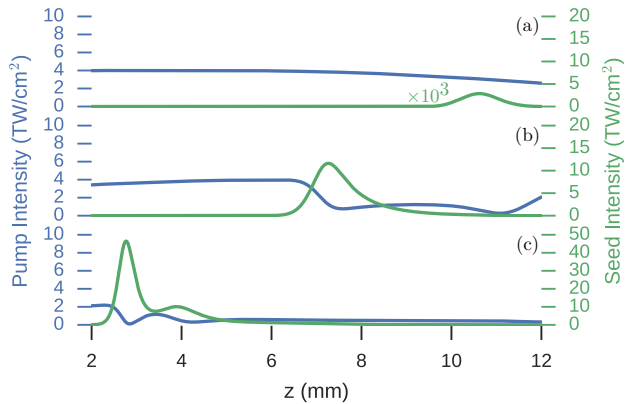}
        \caption{ 
        The pump (blue) and seed (green) intensity at three times.  
        (a) At $t=\SI{97}{\pico\s}$, the seed has entered the plasma and is undergoing exponential growth.  
        (b) At $t=\SI{110}{\pico\s}$, the seed has become sufficiently intense to almost fully deplete the pump.
        (c) At $t=\SI{123}{\pico\s}$, the seed has continued depleting the pump while temporally compressing.
        The seed will continue to grow after the final frame and will reach a final intensity and FWHM duration of \SI{54}{\tera\watt\per\square\cm} and \SI{1.3}{\pico\s}, respectively. 
        The plasma density is \SI{1e17}{\per\cubic\cm}.
        The seed intensity of (a) is shown with an amplitude increased by          a factor of $10^3$.    
        \label{fig: time series}}
\end{figure}

This process is illustrated by the simulation results of FIG.~\ref{fig: time series}.
The pump pulse (blue) is moving in the forward direction (left-to-right) while the seed pulse (green) is moving backwards (right-to-left). 
A uniform plasma covers the entire region. 
Both, the pump and seed are linearly polarized. 
The seed pulse is injected once the pump pulse is overlapping the plasma, as shown in FIG.~\ref{fig: time series}(a).
The initial seed pulse has a full width at half maximum (FWHM) duration of \SI{3}{\pico\s} and intensity of \SI{1e8}{\watt\per\cm\squared}. 
The seed pulse grows exponentially at the small-signal gain rate for Raman backscattering $\gamma_0 = a_0 \sqrt{\omega_0 \omega_p} / 4 = \SI{2.20}{} ( \lambda_0 [\SI{}{\micro\m}] I_0 [\SI{}{\watt\per\cm\squared}])^{1/2} (n_e [\SI{}{\per\cubic\cm}])^{1/4} \SI{}{Hz}$, where $a_0$ is the pump normalized vector potential, 
$\omega_0$ is the pump frequency, 
$\omega_p$ is the plasma frequency, 
$\lambda_0$ is the pump wave length, 
$I_0$ is the pump intensity, and 
$n_e$ is the plasma density \cite{kruer1988physics}.
The results of exponential growth can be seen in FIG.~\ref{fig: time series}(a) as the seed intensity increases by a factor of $30$ after entering the plasma with a peak intensity of \SI{1e8}{\watt\per\cm\squared}.
This corresponds to an approximate gain rate of $\gamma \sim 1/ \SI{3.7}{\pico\s}$.
During the exponential growth regime, the seed pulse duration will lengthen, reaching a maximum duration of \SI{5.2}{\pico\s} in this example. 
This is the result of enhanced backscatter due to a build up in the plasma wave at the tail of the seed pulse. 
Figure \ref{fig: time series}(b) shows when the seed becomes sufficiently intense to deplete the pump. 
At this point, the seed is operating in the pump-depletion regime \cite{malkin1999fast}.
The leading edge of the seed is backscattering enough of the pump that it shadows the trailing edge of the seed. 
This results in temporal gain compression and can be seen in FIG.~\ref{fig: time series}(b) and \ref{fig: time series}(c), with seed durations of \SI{3.3}{\pico\s} and \SI{1.5}{\pico\s}, respectively.   
Figure \ref{fig: time series}(c) shows that pump depletion continues.
The seed intensity grows roughly linearly with propagation distance as it sweeps up the pump energy. 

\iftoggle{memo}{ 
Researchers at Princeton proposed a plasma-based approach for pulse compression that takes advantage of a nonlinear regime of amplification, pump-depletion,  where the pulse would shorten as it grew and it's intensity would be proportional to the square of the number of emitters (plasma electrons) \cite{shvets1998}. 
The formalism for this process was originally developed for growth from noise \cite{shvets1997analysis}. 
To access this regime, a pulse must be intense enough $ \omega_B \ge \omega_p$ and short enough $\tau \approx \pi/ \omega_B$, where 
$\omega_B=2 \omega_0 \sqrt{a_0 a_1}$ is the natural frequency of an electron in a plasma wave (also called the bounce frequency),
$\omega_0$ is the pump frequency, 
$\tau$ is the pulse duration, 
$a_{0,1}$ are the normalized pump and seed vector potentials, 
$e$ is the electron charge, 
$m$ is the electron mass, and
$c$ is the speed of light \cite{kruer1988physics}.

An early proposal suggested the possibility of backward Raman amplification for generating exawatt laser pulses  \cite{malkin1999fast} which would open access to new fundamental physical regimes \cite{tajima2002zettawatt}.
In order for chirped pulse amplification to reach higher power, excessively large gratings are required to keep the laser fluence below the grating damage threshold, typically on the order of \SI{0.1}{\joule\per\cm\squared}.
Plasma-based amplification avoids this issue by not using gratings to stretch the pulse.
Additionally,  material damage thresholds of \SI{}{\giga\watt\per\cm\squared} are avoided by the use of a plasma instead of a crystal as a gain medium. 
}{} 

\iftoggle{memo}{ 
The critical power for relativistic self-focusing of a plasma is $P_\text{cr} \approx 17 (\omega_0/\omega_p)^2 \SI{}{\giga\watt}$. 
This is the power at which plasma nonlinearities, and instabilities, begin to become significant. 
By having the amplification occur faster than all other instabilities, pulses above the critical power could be generated. 
The development of the pi-pulse model dominates the theory of backward Raman amplification \cite{malkin1999fast}. 
This uses a three-wave model to describe the pump, seed, and plasma wave. 
The effects of modulation instability (filamentation), wave breaking, and forward Raman scattering on amplification have been discussed at length. 
It was suggested that the ideal place to operate is near wave breaking and with the highest pump energy that avoids filamentation. 
The pi-pulse amplitude and energy should grow linearly with propagating distance. 
Simultaneously, the pulse duration compresses inversely proportional to propagation distance.  
}{} 

There is an existing body of literature on plasma-based backward Raman amplification. 
Three dimensional particle-in-cell simulations have shown \SI{}{\peta\watt} power pulses being generated by backward Raman amplification at wavelengths of \SI{700}{\nano\meter} and \SI{10}{\nano\meter} \cite{trines2011} but subsequent work suggests this is an overestimate \cite{toroker2014backward,edwards2015efficiency}.
Experimentally, the observed peak output powers are in the range of \SI{60}{\giga\watt} with efficiencies of \SI{6.4}{\percent} for a pump wavelength of \SI{800}{\nano\m} \cite{ren2007new,ren2008compact}.
Previous work \cite{malkin2000ultra} mentioned the possibility of scaling to the \SI{10}{\micro\m} wavelength but for parameters where the initial seed intensity was equal to the pump intensity. 
While this is feasible for near-infrared or visible wavelengths, it is not for LWIR. 
This makes the exponential growth regime critical to the final seed output.

\section{Design Considerations for the Long-wavelength Infrared Regime}

Without high power seed sources at wavelengths of \SI{10}{\micro\m}, the seed must make use of both the small signal gain regime and depletion regime for backward Raman amplification to be effective. 
The small signal gain regime is required in order to get the seed into the depletion regime, at which point a significant fraction of the pump energy can be transfered to the seed pulse.

\begin{figure}[ht!] 
    \includegraphics[width=0.49\textwidth]{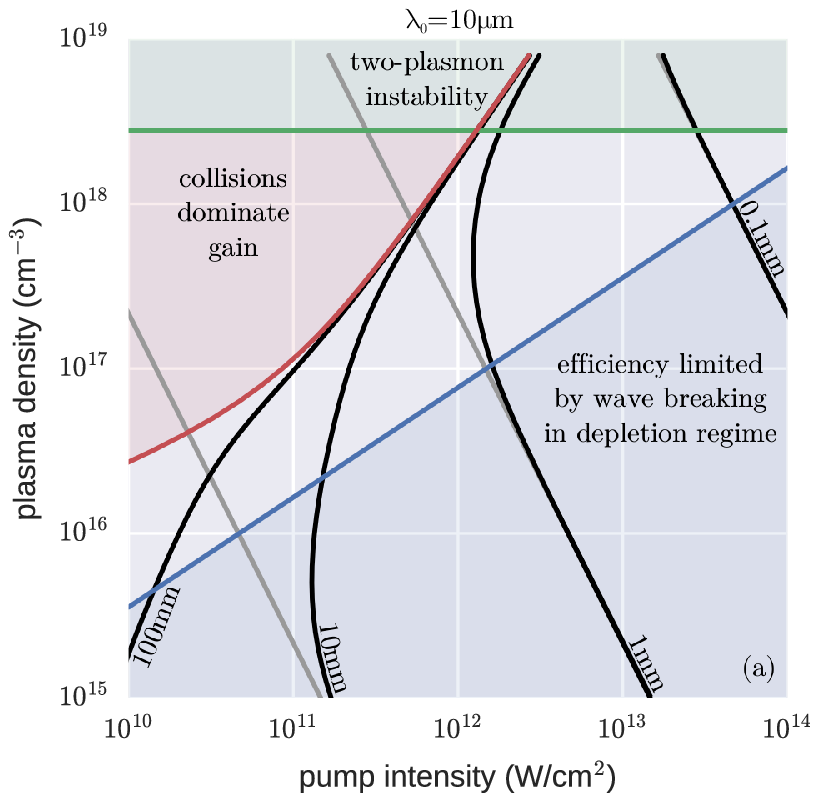}
    \includegraphics[width=0.49\textwidth]{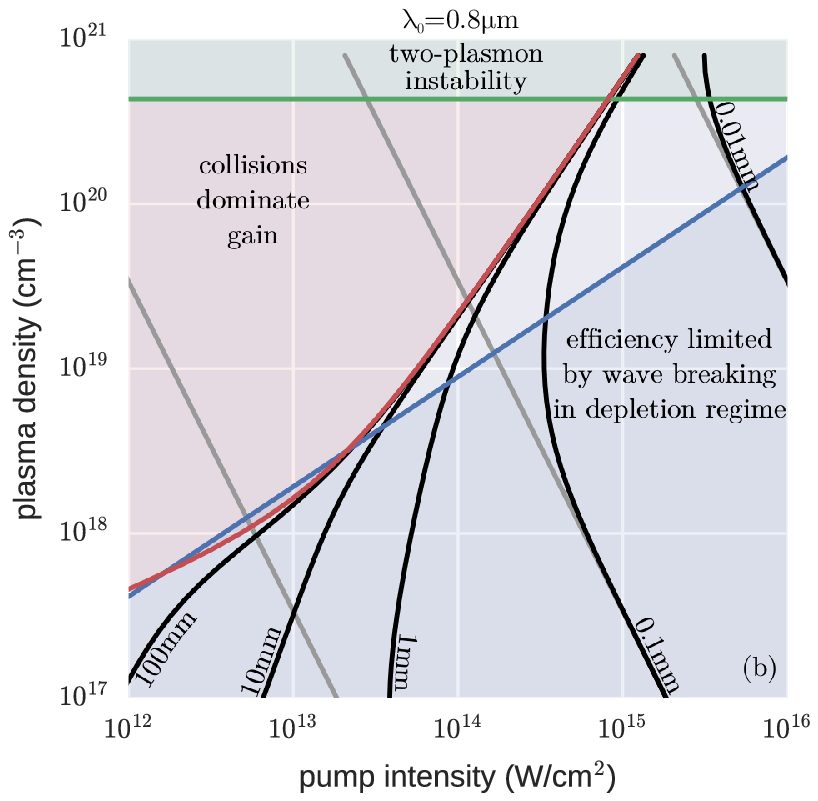}  
    \caption{
            Contours (black) of the steady-state Raman backscatter gain 
            length,$c/\gamma$, are shown for 
            (a) $\lambda_0=\SI{10}{\micro\m}$ and 
            (b) $\lambda_0=\SI{0.8}{\micro\m}$.
            The contours of the gain length in a collisionless plasma are shown
            by the gray lines. 
            The red region shows where there is no gain due to collisions 
            $\gamma_0^2<\nu_e\nu_s/4$.  
            The plasma is singly ionized with temperature $T=\SI{1}{\eV}$,
            $\ln \Lambda =10$, $Z=1$.
            The blue region shows where wave breaking would reduce the 
            efficiency once the seed is in the depletion regime. 
            The green region shows when two-plasmon decay would deplete the pump.
    \label{fig: rbs gain rate both figs}}
\end{figure}

The pump intensity and plasma density are major design considerations as both directly influence the gain rate, wave breaking, and other instabilities.
Figure \ref{fig: rbs gain rate both figs}(a) shows a contour plot illustrating various design limitations in the plasma density and pump intensity parameter space. 
The gray lines show the time-independent backward Raman gain length, the length over which the seed field strength will increase by a factor of $e^1$, given by $c / \gamma_0$, 
where $\gamma_0 = \sqrt{\omega_0 \omega_p} a_0 / 4$ is the collisionless gain rate.
The black lines show the effect of electron-ion collisions on the gain length $c/\gamma$, 
where $\gamma$ is the collisional gain rate given by Eq.~\eqref{equ: gain rate}  \cite{kruer1988physics,strozzi2005vlasov}. 
The electron-ion collision rate depends on the plasma temperature and ponderomotive energy with a functional form approximated by $\nu_{e} \sim \left( T + U_p \right)^{-3/2}$.
While time-dependent effects are important \cite{malkin1999fast}, the time-independent gain rate is illustrative for understanding the trade-offs in pump intensity and plasma density.

Shorter gain lengths are produced with higher pump intensities and plasma densities because both enhance the plasma wave.
At high plasma density, electron-ion collisions ($\nu_{e}\propto n_e$) will damp out the plasma wave faster than growth from backward Raman amplification ($\gamma_0\propto n_e^{1/4}$) \cite{kruer1988physics,strozzi2005vlasov}.
The difference in the collisional and collisionless gain lengths (black and gray curves of FIG.~\ref{fig: rbs gain rate both figs}(a)) show where collisional damping becomes important.
The red lines mark the instability threshold where the gain is zero due to collisional damping and is given by $\gamma_0^2 = \nu_e\nu_s/4$, where $\nu_s$ is the energy damping rate of the electromagnetic wave (see Appendix \ref{sec: gain rate}) \cite{kruer1988physics}.  
The isocontours of the collisional gain length (black curve) compress against the boundary where collisions dominate gain (red curve) because as the gain approaches zero, the gain length goes to infinity. 
Additionally, the collisional damping is suppressed at pump intensities where the ponderomotive energy is greater than the plasma temperature $I_0> \SI{1.07e13}{\watt\per\cm\squared} T[\SI{}{\eV}] / (\lambda_0 [\SI{}{\micro\m}])^2$. 
This can be observed in FIG.~\ref{fig: rbs gain rate both figs}(a) where the red line changes slope. 

The green line at quarter critical density $n_\text{crit}/4 \approx \SI{2.8e20}{\per\cubic\cm}/ (\lambda[\SI{}{\micro\m}])^{2}$ marks where the two-plasmon decay occurs and where the Raman instability becomes absolute. 
This sets a hard upper bound on the possible plasma density, otherwise the pump would be rapidly absorbed by the plasma. 
Typically, other instabilities (forward Raman, parasitic backward Raman, or filamentation) will dominate at plasma densities below quarter critical \cite{trines2011}.

Plasma wave breaking can limit efficient depletion of the pump.
During depletion, each pump photon is stimulated to scatter into an additional seed photon and plasmon.
The plasma wave, however, has a maximum energy density that it can support \cite{kruer1988physics}, in other words, a maximum plasmon density.
Therefore, when the maximum plasmon density has been reached, the seed growth is stunted.
The blue curve of FIG.~\ref{fig: rbs gain rate both figs}(a) marks the largest pump intensity that the plasma can support during 100\% depletion \cite{malkin2014key}.
The pump intensity should be below $I_0 = \SI{4.7e-15}{} \lambda_0 [\SI{}{\micro\m}]\left( n_e [\SI{}{\per\cubic\cm}]\right)^{3/2} \SI{}{\watt\per\cm\squared}$ \cite{malkin2014key} to avoid wave breaking.
Plasma wave breaking does not affect the small signal growth of the seed pulse, but it will change the overall efficiency of the growth when the seed is in the depletion regime \cite{malkin1999fast}.
Therefore this is a soft limitation and there is evidence that some wave breaking is preferable as it can limit energy growth in the tail of the seed pulse\cite{malkin1999fast}.

Within these constraints, we can see the gain length varies from \SI{100}{\milli\m} to \SI{0.1}{\milli\m}.
There are, however, experimental limitations to realizing a particular plasma length, density, and uniformity. 
In particular, available sources generate long-wavelength infrared pulses with microjoules of energy, requiring four to five e-foldings in the linear growth regime.
The need of using both the linear and nonlinear regimes is distinct from previous studies which focused primarily on near-infrared or visible light where intense, short pulses for seeding are readily available.

The parameter space for \SI{10}{\micro\m} can be contrasted with FIG.~\ref{fig: rbs gain rate both figs}(b)  which shows the gain length as a function of pump intensity and plasma density for \SI{800}{\nano\m}. 
It is clear that the operating window is smaller for \SI{800}{\nano\m}, but this is less significant because there are not the same technological limitations on seed pulse generation.

In summary, this  motivates using laser pump intensities of around \SI{1e12}{\watt\per\cm\squared} and plasma densities around \SI{1e17}{\per\cubic\cm} in order to avoid collisional damping, wave breaking, and have a sufficiently small gain length. 
We note that the conculsions hold for an electron temperature of \SI{1}{\eV}.
A lower pump intensity is possible for higher electron temperatures. 
\section{Numerical Results}

Motivated by the regime described above, we carried out one and two-dimensional simulations using the turboWAVE framework, which couples a finite-difference time-domain (FDTD) electromagnetic solver with a collisional,  relativistic cold fluid plasma model \cite{gordon2000}.  
The simulation domain consists of several sections in the following order; a ``vacuum'' section, an up ramp, a uniform plasma section, a down ramp, and a final ``vacuum'' section. 
The uniform plasma section has a density $n_0$ and length $L_z$.
The ``vacuum'' sections have an electron density $10^{-4} n_0$.
The ``vacuum'' and ramp sections are all \SI{1}{\milli\m} long. 
The electron-ion collision model requires a constant plasma temperature which is set at \SI{1}{\eV}. 

\begin{figure}[!ht]
    \includegraphics[width=\figurewidth]{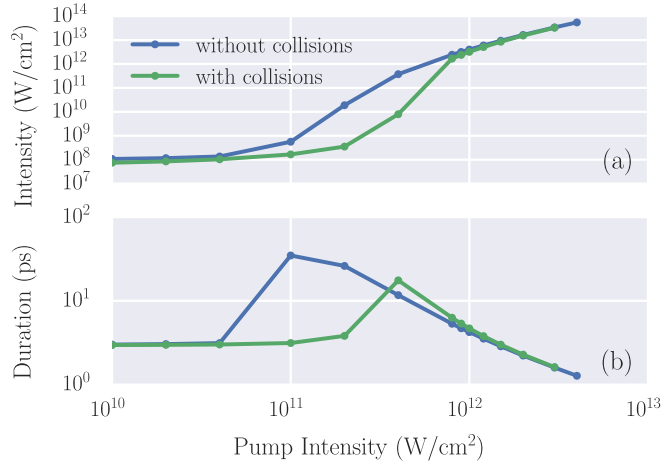}
    \caption{ The final seed (a) peak intensity and (b) FWHM duration are shown as a function of pump intensity. 
    The blue (green) curves show the result of neglecting (including) the effect of electron-ion collisions. 
    The maximum seed intensity reached is \SI{5.5e13}{\watt\per\cm\squared} with a duration of \SI{1.3}{\pico\s} for a pump intensity of \SI{4e12}{\watt\per\cm\squared}. 
    \label{fig: pump intensity scan with and without collisions}}
\end{figure}

A scaling of the pump intensity for fixed plasma density was carried out to illustrate the effects of collisional damping, the exponential gain regime, and the depletion regime on final intensity and FWHM duration of the seed pulse.
These can be seen in FIG.~\ref{fig: pump intensity scan with and without collisions}.
For this specific set of simulations, 
the plasma density is $n_0 = \SI{1e17}{\per\cubic\cm}$, 
the plasma frequency is  $\omega_p = \SI{1.78e13}{\radian\per\s}$, 
the grid size is $0.01k_p^{-1}$, and
the time step is $0.009\omega_p^{-1}$.

The pump pulse ($\lambda_0 = 10.6\mu\text{m}$) enters the left side of the simulation domain at $t=0$.
The initial seed pulse ($\lambda_1 = 11.8\mu\text{m}$) enters the right side of the constant plasma density region as the pump pulse at the left side reaches half its peak intensity.
The seed pulse frequency was chosen to be on resonance for the Raman instability $\omega_1 = \omega_0 - \omega_p$. 
The initial seed pulse's FWHM duration is \SI{3}{\pico\s} with \SI{300}{\micro\joule\per\square\cm} contained within.
The initial fluence was chosen so that with a \SI{1}{\milli\m\squared} cross-section, the initial seed pulse would have \SI{}{\micro\joule} energies. 
The plasma length is $L_z = \SI{10}{\milli\m}$.
The pump pulse FWHM duration is \SI{68}{\pico\s}, which corresponds to twice the plasma length $2 L_z/c$. 
Field values were recorded at the simulation boundaries.
Spectral box filters from $\pm\omega_p/2$ around the pump and seed frequencies were used to extract individual field envelopes and intensity profiles.

Figure \ref{fig: pump intensity scan with and without collisions} shows the dependence of the amplified seed on pump intensity with and without electron-ion collisions, the green and blue curves, respectively.
At intensities below \SI{1e11}{\watt\per\cm\squared}, there is no significant gain in the seed pulse because the gain length is comparable to the plasma length, as seen in FIG.~\ref{fig: rbs gain rate both figs}(a). 
As the pump intensity is increased from \SI{1e11}{\watt\per\cm\squared} to just below \SI{1e12}{\watt\per\cm\squared}, the seed intensity in the collisionless plasma grows exponentially with pump intensity. 
This means that an increase in the pump intensity allows the seed to undergo additional e-foldings within the fixed plasma length. 
This stops at just below \SI{1e12}{\watt\per\cm\squared}, where the seed begins growing more slowly with pump intensity. 
The seed pulse has grown sufficiently intense that it is beginning to deplete the pump.
Similar to what occurs in FIG.~\ref{fig: time series}, once the seed pulse begins depleting the pump, the rate of increase in the seed intensity is no longer exponential, but roughly linear with the encountered pump fluence. 
For fixed plasma length, this suggests that the seed will grow roughly linearly with pump intensity. 
The rate of increase in FIG.~\ref{fig: pump intensity scan with and without collisions} is faster than linear, because a higher pump intensity shortens the length needed for the seed to exponentiate and reach depletion.
This increases the length over which the seed can deplete and, hence, the seed grows faster than linear with pump intensity. 
For a pump intensity of \SI{4e12}{\watt\per\cm\squared}, the maximum seed intensity reached \SI{5.5e13}{\watt\per\cm\squared} with a duration of \SI{1.3}{\pico\s}. 
For a \SI{1}{\milli\m} cross-sectional area, this corresponds to a peak power of \SI{0.55}{\tera\watt}, energy of \SI{0.72}{\joule}, and amplification factor of \SI{2.4e5}{}.
Further increase in pump intensity may be beneficial, but cannot be simulated using a fluid model due to wave breaking.

The temporal dynamics of the seed pulse also show the exponential growth and depletion regimes. 
When the gain length is longer than the plasma length, the seed's final duration is equal to the initial duration, as seen in the first three points of FIG.~\ref{fig: pump intensity scan with and without collisions}(b).
During the small signal gain regime, the point of maximal growth sweeps backwards at $c/2$ \cite{malkin2000detuned}.
Essentially, the leading edge of the seed is driving a plasma wave causing increased backscatter later in the seed pulse. 
This can be seen most clearly in the collisionless plasma where seed durations of \SI{32}{\pico\s} are observed.
When collisions are included, the seed duration still grows during the linear regime but only at higher pump intensities where the electron-ion collision rate is less important.  
Once the seed begins depleting the pump, the seed duration rapidly decreases.
What is occurring is not compression of the seed energy, but amplification in the seed's leading edge, namely temporal gain compression. 
Pulse durations as short as \SI{1.3}{\pico\s} are reached. 
This is shorter than the \SI{2}{\pico\s}-\SI{3}{\pico\s} pulses that are typically created in high power \ce{CO2} lasers \cite{pigeon2014generation,pogorelsky2015bestia}. 
Previous work has shown that some degree of wave breaking is beneficial.
When wave breaking occurs after the peak of the seed pulse has passed, it can suppress the growth in trailing pulses, such as those seen in FIG.~\ref{fig: time series}(c) \cite{malkin1999fast}.
   
\begin{figure}[!ht]
    \includegraphics[width=\figurewidth]{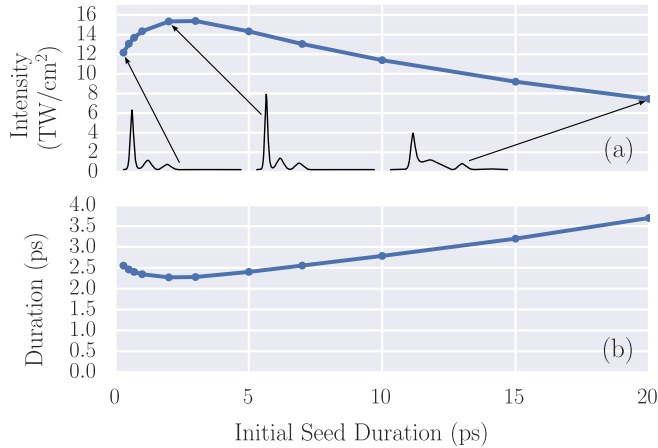}
    \caption{ The final seed (a) peak intensity and (b) FWHM duration are shown as a function of the initial seed duration. 
    These are fluid turboWAVE simulations with electron-ion collisions. 
    The pump intensity is \SI{2}{\tera\watt\per\cm\squared}, 
    plasma density is \SI{1e17}{\per\cubic\cm},
    and plasma length is \SI{10}{\milli\m}. 
    The black lineouts show the final intensity profiles of the seed pulse. 
    The lineout axes cover a temporal and intensity range of 
    \SI{80}{\pico\s} and \SI{16}{\tera\watt\per\cm\squared}. 
    \label{fig: seed duration scan with collisions}}
\end{figure}

Figure \ref{fig: seed duration scan with collisions} plots the final seed intensity and pulse duration as a function of the initial seed duration, for a fixed fluence. 
The pump intensity is \SI{2e12}{\watt\per\cm\squared} and all other parameters are the same as FIG.~\ref{fig: pump intensity scan with and without collisions}.
There is a weak dependence of the seed's final intensity and duration on its initial duration. 
This suggests that the initial seed pulse duration is not a critical design parameter.

The intensity maximum in FIG.~\ref{fig: seed duration scan with collisions}(a) can be understood from the dynamics of the small signal gain.
The point of maximal gain sweeps backward at half the speed of light \cite{malkin2000detuned}. 
If the seed it too short, the point of maximal gain will pass over it and amplify its weak tail. 
This can seen in the leftmost inset intensity profile of FIG.~\ref{fig: seed duration scan with collisions}(a).
If the seed is too long, the head of the seed will amplify to the depletion regime before the tail.
In this case, the pulse would have reached depletion sooner had the pulse been shorter.  
This can seen in the rightmost inset intensity profile of FIG.~\ref{fig: seed duration scan with collisions}(a), because energy was transfered to the long, secondary pulse during exponential growth, delaying the onset of pump depletion and reducing the time for the short, primary pulse to grow. 

\begin{figure}[!ht]
    \includegraphics[width=\figurewidth]{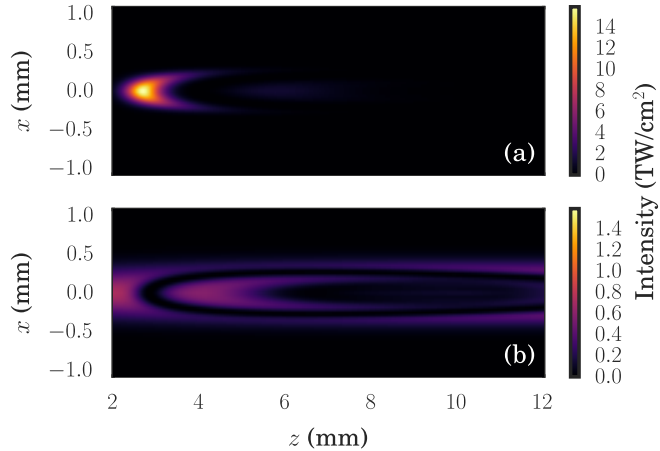}
    \caption{Intensity of seed (a) and pump (b) just before the seed exits the plasma region. 
    \label{fig: 2d pulse intensities}}
\end{figure}

Two-dimensional turboWAVE simulation results are shown in FIG.~\ref{fig: 2d pulse intensities}.
This simulation is similar to those previously shown with several differences. 
The initial seed intensity and duration are \SI{1e10}{\watt\per\cm\squared} and \SI{3}{\pico\s}.
The pump and seed's initial $e^{-1}$ field spot sizes are \SI{0.4}{\milli\m}.
The pump intensity is \SI{7.4e11}{\watt\per\cm\squared}. 
The geometry is planar.
The longitudinal and transverse spatial coordinates are $z$ and $x$, respectively.  
Along the longitudinal direction, the \SI{10}{\milli\m} plasma is constructed the same as the one-dimensional simulations. 
The plasma profile in the transverse direction is uniform with periodic boundary conditions.
The simulation domain had 83328 and 64 cells  in the z- and x-directions, respectively. 
The cell size is $\Delta z = 0.01k_p^{-1}$ and $\Delta x = 2k_p^{-1}$ in the z- and x-directions, respectively.  

The depletion of the pump can be seen in FIG.~\ref{fig: 2d pulse intensities}(b).
The seed intensity has increased by a factor of 360. 
The seed has a smaller spot size due to gain focusing.  
This shows that comparable results are possible in one- and two-dimensional simulations. 
The transverse grid size is too large to resolve plasma perturbations needed to drive the filamentation instability. 
Further investigation is needed to determine the importance of that effect.

\section{Numerical Limitations}

Two modeling limitations have been observed when simulating backward Raman amplification. 
First, the fluid simulations are limited by wave breaking of the plasma which occurs for pump intensities around \SI{2e12}{\watt\per\cm\squared} at wavelengths of \SI{10.6}{\micro\m} and densities of \SI{1e17}{\per\cubic\cm}.
The most promising cases for amplification tend to occur when the pump intensity is at or above the wave breaking limit, as suggested by FIG.~\ref{fig: rbs gain rate both figs}.
Specifically, this occurred for pump intensities of \SI{4e12}{\watt\per\cm\squared}.
The pump pulse alone will not drive a significant plasma wave by itself.
As the seed is amplified, it beats with the pump and drives the plasma wave to break.
Fundamentally, this limits the maximum seed intensities that are possible to reach in fluid simulations at a given pump intensity.  

Particle-in-cell (PIC) simulations offer the ability to model the amplification process in the regime above plasma wave breaking.  
Extensive PIC simulations have been carried out and show quantitative agreement with the collisionless fluid model. 
However, when the number of potential e-foldings is sufficiently large, $\gamma_0 L_z/c \sim 4$ to $5$, parasitic Raman backscatter competes with the amplification process and limits growth. 
The seeding of the parasitic process is several orders of magnitude larger in the PIC versus fluid simulations. 
This limits the use of a PIC plasma model for this problem because it is of practical importance to start with an initially weak seed pulse.  
Detuning the interaction with a spatially varying plasma density \cite{malkin2000detuned} may suppress amplification of noise and make PIC simulations for these parameters more feasible, 
but the need of multiple e-foldings for the seed to reach depletion makes the detuning technique more challenging.  

\section{Conclusion}

Simulations have demonstrated that backward Raman amplification can compress and amplify LWIR pulses. 
The turboWAVE framework has been used to carry out one- and two-dimensional FDTD electromagnetic simulations coupled to a relativistic cold fluid plasma model with electron-ion collisions. 
Using a pump pulse that could be generated by a \ce{CO2} laser, it was shown that a seed pulse at \SI{11.7}{\micro\m} could be amplified to \SI{5.4e13}{\watt\per\cm\squared} and compressed to a duration of \SI{1.3}{\pico\s}.
When compared to the initial pump pulse, the final seed pulse is 10 times more intense and 50 times shorter. 
The final seed amplification is weakly dependent on the initial seed duration, which is promising as sources are limited.

Limitations in available long wavelength infrared sources motivated the use of both the linear and nonlinear growth regimes.  
For large plasma densities, collisional damping can eliminate growth unless pump intensities are sufficiently intense to compensate for, or suppress, damping.
This is particularly important during the linear growth regime. 
In the depletion regime, plasma wave breaking provides a soft upper limit on the pump intensity by limiting the depletion efficiency.
At large plasma densities and pump intensities, two-plasmon decay or absolute Raman will deplete the pump, but a more detailed analysis of the other limiting instabilities is needed. 
The ultimate limits on efficiency could not be determined because of numerical difficulties.  
Future work should include a collisional kinetic model in which the noise source can be controlled and a study of the importance of plasma length and temperature.

\appendix
\section{Temporal Weak Coupling Gain}\label{sec: gain rate}

As derived elsewhere \cite{strozzi2005vlasov,kruer1988physics}, the Raman backscattering dispersion relation for the plasma density perturbations with frequency $\omega$ and wavenumber $\vec{k}$ is 
\begin{equation}
\left[ \omega^2 -\omega_{ek}^2 + i \nu_{e} \omega \right]
\left[ \left(\omega - \omega_0\right)^2 - \left(\vec{k}-\vec{k}_0\right)^2 c^2 
-\omega_{p}^2 + i \nu_{s} \left( \omega - \omega_0\right) \right] = 
\frac{\omega_{p}^2 k^2 v_\text{os}^2 }{4},
\end{equation}
where the Bohm-Gross frequency is $\omega_{ek}^2 = \omega_{p}^2 + 3 k^2 v_\text{th}^2$,
the electron thermal velocity is $v_\text{th} = \sqrt{T/m}$,
the electron temperature is $T$, 
the energy damping rate of the electron plasma wave $\nu_{e}$,
the pump frequency and wavenumber are $\omega_0, \vec{k}_0$, 
the energy damping rate of the electromagnetic wave (inverse-bremsstrahlung) is $\nu_s \approx (\omega_p/\omega_0)^2  \nu_{e}$,
and the quiver velocity is $v_\text{os} = e A_0 / (m c)$,
where the pump vector potential is 
$\vec{A}_L = \hat{x} A_0 \cos (k_0 z - \omega_0 t )$ \cite{kruer1988physics}. 

The scattered electromagnetic wave should have a frequency $\omega_s \approx \omega_0 - \omega_p$. 
A frequency detuning of $\Delta \omega$ from the resonant frequency $\omega_0-\omega_{ek}$ defines the scattered wave frequency $\omega_s = \omega_0-\omega_{ek} + \Delta\omega$. 
The plasma wave should have approximately the Bohm-Gross frequency.
We will define the frequency with a real frequency shift of $\Delta\omega$ for detuning and a complex shift of $\delta\omega$ for gain and collisional damping, that is, 
$\omega = \omega_{ek}-\Delta\omega+\delta\omega$.
After making the approximation that $\delta\omega,\Delta\omega \ll \omega_{ek}$ the dispersion relation reduces to 
\begin{equation}
\left( \delta\omega + i \nu_s/2\right) 
\left( \delta\omega  + i \nu_e/2 -\Delta\omega \right) 
= - \gamma_0^2,
\end{equation} 
where the resonant, collisionless gain rate is 
\begin{equation}\label{equ: collisionless gain rate}
\gamma_0^2 = \frac{\omega_{p}^2 k^2 v_\text{os}^2 }{16 \omega_{ek} (\omega_0-\omega_{ek})}.
\end{equation}
The complex frequency shift of the plasma density perturbation is 
\begin{equation}\label{equ: gain rate}
\delta\omega = 
- i \frac{\nu_e}{2} 
+ i \left[ \frac{\nu_e-\nu_s}{4} - i \frac{\Delta\omega}{2} 
+ \sqrt{\gamma_0^2 + \left( \frac{\nu_e-\nu_s}{4} - i \frac{\Delta\omega}{2} \right)^2 } \right].
\end{equation}

In the limit of no collisions $\nu_e,\nu_s\rightarrow 0$ and resonant Raman $\Delta\omega\rightarrow 0$,  $\delta\omega = i\gamma_0$. 
In the limit of resonant Raman $\Delta\omega \rightarrow 0$ but no Raman gain $\gamma_0 \rightarrow0$, then the frequency shift just accounts for the collisional damping of the plasma wave $\delta\omega = - i \nu_e/2$.
In the limit of no Raman gain $\gamma_0 \rightarrow 0$ and no collisions $\nu_e,\nu_s \rightarrow 0$, there is no induced frequency shift $\delta\omega = 0$.

If the interaction is resonant, $\Delta\omega =0$, then we get the following condition for an instability $\gamma_0^2 \ge \nu_e \nu_s / 4$ \cite{kruer1988physics}.
If $\nu_e=\nu_s=0$, then $\delta \omega = \Delta\omega /2 +  i \sqrt{ \gamma_0^2 - \Delta\omega^2/4}$ which says that an instability only exists for $\Delta\omega < 2\gamma_0$.
The power spectrum after an interaction time of $T$, will be proportional to $\exp[2\sqrt{\gamma_0^2 -\Delta\omega^2/4}T]$. 
In the limit that the detuning is much less than the resonant, collisionless gain rate, the FWHM duration of the seed intensity will be approximately $\tau_\text{FWHM} = \sqrt{\ln{2} T/\gamma_0}$. 
This suggests that during the linear growth regime, the pulse durations will grow with time due to gain narrowing, and shorten with increased gain rate due to a larger gain bandwidth.

\section{Collisional Damping} \label{sec: collision rate}

Collisional damping can play a significant role in backward Raman amplification depending on the pump intensity and plasma density. 

The intensity-dependent electron-ion collision rate is given by 
$\nu_{e} = 3 \nu_0 (v_\text{os}/2v_\text{th})^{-3}$ $Q(v_\text{os}/2v_\text{th})$, 
 where the zero intensity rate is $\nu_0 = 4\sqrt{2\pi} Z^2 e^4 n_e \ln{\Lambda} / (3 m^{1/2} T^{3/2} ) \approx \SI{2.91e-5}{}$ $n_e[\SI{}{\per\cubic\cm}] (T[\SI{}{\eV}])^{-3/2} \SI{}{\Hz}$,
 the quiver velocity is $v_\text{os} \approx 25.7 \lambda [\SI{}{\micro\m}] \sqrt{ I[\SI{}{\watt\per\square\cm}] } \SI{}{\cm\per\s}$, 
%
%
the thermal velocity is $v_\text{th} \approx \SI{4.19e7}{} \sqrt{T[\SI{}{\eV}]} \SI{}{\cm\per\s}$,  
 $Q(x)=\int_{0}^{x} dz z^2 \left[ I_0 \left( z^2 \right) - I_1 \left( z^2 \right) \right]e^{-z^2}$, and
the functions $I_n(z)$ are modified Bessel functions of the first kind. 
The electron charge and mass are $e$ and $m$. 
The Coulomb logarithm is $\ln \Lambda =10$ and ionization degree is $Z=1$. 
The electron-ion collision rate is proportional to the plasma density and inversely proportional to the electron velocity cubed. 
The thermal and quiver velocity play a role in the overall collision rate, as can be seen by $\nu_0$ and $3(v_\text{os}/2v_\text{th})^{-3} Q(v_\text{os}/2v_\text{th})$, respectively.
This has been investigated in detail \cite{catto1977strong}. 
As laser intensity increases and the quiver velocity becomes greater than the thermal velocity, the collision rate begins to decrease. 
\begin{figure}[!ht]
    \includegraphics[width=\figurewidth]{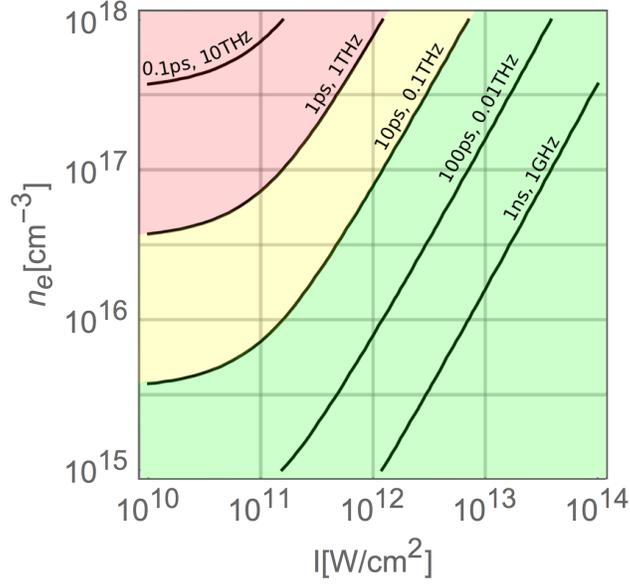}
    \caption{ The electron-ion collision time for $T=\SI{1}{\eV}$,  
    $\lambda = \SI{10}{\micro\m}$, $\ln \Lambda =10$ $Z=1$. 
        The green, yellow, and red regions show parameter regimes where 
        the collision time is good, mediocre, and poor. 
    \label{fig: ei collision time}}
\end{figure}

The effect of laser polarization on the electron-ion collision rate was estimated to not cause a difference larger than a factor of $3$ in the rate \cite{catto1977strong}.

\begin{figure}[!ht]
    \centering
            \includegraphics[width=\figurewidth]{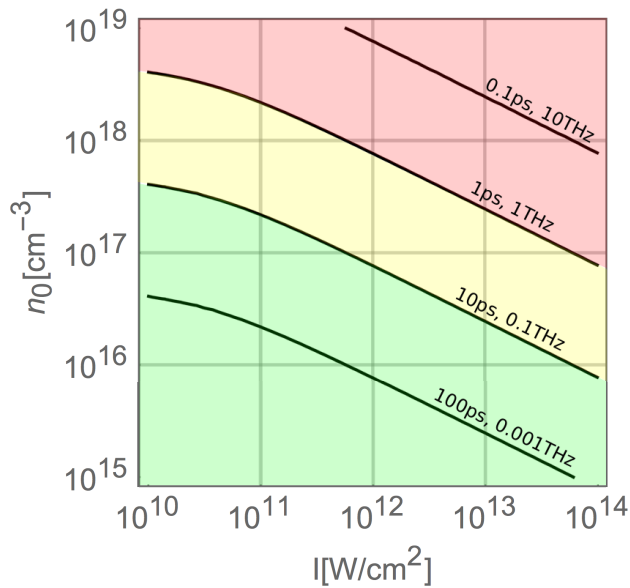}
            \caption{ The electron-neutral collision time for $T=\SI{1}{\eV}$,
            $\lambda = \SI{10}{\micro\m}$, cross-section $\sigma_{en}=\SI{5e-15}{\square\cm}$. 
            The green, yellow, and red regions show parameter regimes where 
            the collision time is good, mediocre, and poor relative to a \SI{1}{\pico\s} pulse duration. 
            \label{fig: en collision time}}
    \end{figure}

Electron-neutral collisions are not considered in the simulations but are a significant consideration if the plasma is not fully ionized. 
The collision rate is given by $\nu_{en} = n_0 \sigma_{en} v_\text{th}$ where $n_0$ is the neutral density, $\sigma_{en} \approx \SI{5e-15}{\square\cm}$ \cite{huba2004nrl}.
To include the increased rate of collisions due to electron quiver, the following substitution can be used  $v_\text{th} \rightarrow \sqrt{v_\text{th}^2 + v_\text{os}^2}$.
An approximate expression for the electron neutral collision rate is $\nu_{en} \approx \SI{2.10e-7}{} n_0[\SI{}{\per\cubic\cm}] \sqrt{ T[\SI{}{\eV}] + \SI{3.76e-13}{}  (\lambda[\SI{}{\micro\m}])^2 I[\SI{}{\watt\per\cm\squared}]}$.
Figure \ref{fig: en collision time} shows contours of constant electron-neutral collision rate. 
Generation of picosecond duration pulses requires the use of pump intensities and neutral densities where the collisional damping is slower than the seed pulse. 
As seen in FIG.~\ref{fig: en collision time}, this means neutral densities below \SI{1e17}{\per\cubic\cm} would be feasible.

Finally, previous work found evidence of Landau damping \cite{trines2011} but it saturated quickly and was insignificant. 
The damping rate can be estimated by  $\delta \omega / \omega_p = - \frac{i}{2} \sqrt{\frac{\pi}{2}} (k\lambda_D)^{-3}$ $\exp \left[ -\frac{1}{2}(k\lambda_D)^{-2} \right]$ \cite{fitzpatrick2014plasma},  where the characteristic plasma wave wavenumber is $k=2\omega_0/c$ and the Debye length is $\lambda_D = \sqrt{T/(4\pi n_e e^2)}$. 
For a pump wavelength of $\lambda=\SI{10}{\micro\m}$, plasma density of $n_e=\SI{1e17}{\per\cubic\cm}$, temperature of $T=\SI{1}{\eV}$, and  $k\lambda_D = 9.33\times10^7 \sqrt{ T[\SI{}{\eV}] / n_e[\SI{}{\per\cubic\cm}] } / \lambda[\SI{}{\micro\m}] \approx 0.03$, the damping rate relative to the plasma frequency is insignificant. 
At lower densities of \SI{1e15}{\per\cubic\cm}, the relative damping rate is $\text{Im}[\delta\omega]/\omega_p \sim  0.1$ and could warrant further consideration.

The units in this section are cgs-Gaussian unless otherwise stated.  

\begin{acknowledgments}
We would like to acknowledge A. Stamm and N. Fisch for fruitful discussions and J. Rajkowski for careful proofreading. 
Data post-processing and plotting was carried out using the following Python packages; seaborn\cite{seaborn}, pandas \cite{pandas}, numpy\cite{numpy}, and mpmath\cite{mpmath}.  
This work has been supported by the U.S. Naval Research Laboratory's Karle Fellowship.
Resources of the Department of Defense High Performance Computing and Modernization Program (HPCMP) were used in this work.
\end{acknowledgments}

\bibliographystyle{apsrev4-1}
\bibliography{main} 

\begin{thebibliography}{37}%
\makeatletter
\providecommand \@ifxundefined [1]{%
 \@ifx{#1\undefined}
}%
\providecommand \@ifnum [1]{%
 \ifnum #1\expandafter \@firstoftwo
 \else \expandafter \@secondoftwo
 \fi
}%
\providecommand \@ifx [1]{%
 \ifx #1\expandafter \@firstoftwo
 \else \expandafter \@secondoftwo
 \fi
}%
\providecommand \natexlab [1]{#1}%
\providecommand \enquote  [1]{``#1''}%
\providecommand \bibnamefont  [1]{#1}%
\providecommand \bibfnamefont [1]{#1}%
\providecommand \citenamefont [1]{#1}%
\providecommand \href@noop [0]{\@secondoftwo}%
\providecommand \href [0]{\begingroup \@sanitize@url \@href}%
\providecommand \@href[1]{\@@startlink{#1}\@@href}%
\providecommand \@@href[1]{\endgroup#1\@@endlink}%
\providecommand \@sanitize@url [0]{\catcode `\\12\catcode `\$12\catcode
  `\&12\catcode `\#12\catcode `\^12\catcode `\_12\catcode `\%12\relax}%
\providecommand \@@startlink[1]{}%
\providecommand \@@endlink[0]{}%
\providecommand \url  [0]{\begingroup\@sanitize@url \@url }%
\providecommand \@url [1]{\endgroup\@href {#1}{\urlprefix }}%
\providecommand \urlprefix  [0]{URL }%
\providecommand \Eprint [0]{\href }%
\providecommand \doibase [0]{http://dx.doi.org/}%
\providecommand \selectlanguage [0]{\@gobble}%
\providecommand \bibinfo  [0]{\@secondoftwo}%
\providecommand \bibfield  [0]{\@secondoftwo}%
\providecommand \translation [1]{[#1]}%
\providecommand \BibitemOpen [0]{}%
\providecommand \bibitemStop [0]{}%
\providecommand \bibitemNoStop [0]{.\EOS\space}%
\providecommand \EOS [0]{\spacefactor3000\relax}%
\providecommand \BibitemShut  [1]{\csname bibitem#1\endcsname}%
\let\auto@bib@innerbib\@empty
\bibitem [{\citenamefont {Strickland}\ and\ \citenamefont
  {Mourou}(1985)}]{strickland1985compression}%
  \BibitemOpen
  \bibfield  {author} {\bibinfo {author} {\bibfnamefont {D.}~\bibnamefont
  {Strickland}}\ and\ \bibinfo {author} {\bibfnamefont {G.}~\bibnamefont
  {Mourou}},\ }\href {http://dx.doi.org/10.1016/0030-4018(85)90120-8}
  {\bibfield  {journal} {\bibinfo  {journal} {Optics Communications}\ }\textbf
  {\bibinfo {volume} {56}},\ \bibinfo {pages} {219} (\bibinfo {year}
  {1985})}\BibitemShut {NoStop}%
\bibitem [{\citenamefont {Trines}\ \emph {et~al.}(2011)\citenamefont {Trines},
  \citenamefont {Fiuza}, \citenamefont {Bingham}, \citenamefont {Fonseca},
  \citenamefont {Silva}, \citenamefont {Cairns},\ and\ \citenamefont
  {Norreys}}]{trines2011}%
  \BibitemOpen
  \bibfield  {author} {\bibinfo {author} {\bibfnamefont {R.}~\bibnamefont
  {Trines}}, \bibinfo {author} {\bibfnamefont {F.}~\bibnamefont {Fiuza}},
  \bibinfo {author} {\bibfnamefont {R.}~\bibnamefont {Bingham}}, \bibinfo
  {author} {\bibfnamefont {R.}~\bibnamefont {Fonseca}}, \bibinfo {author}
  {\bibfnamefont {L.}~\bibnamefont {Silva}}, \bibinfo {author} {\bibfnamefont
  {R.}~\bibnamefont {Cairns}}, \ and\ \bibinfo {author} {\bibfnamefont
  {P.}~\bibnamefont {Norreys}},\ }\href {https://doi.org/10.1038/NPHYS1793}
  {\bibfield  {journal} {\bibinfo  {journal} {Nature Physics}\ }\textbf
  {\bibinfo {volume} {7}},\ \bibinfo {pages} {87} (\bibinfo {year}
  {2011})}\BibitemShut {NoStop}%
\bibitem [{\citenamefont {Malkin}\ \emph {et~al.}(1999)\citenamefont {Malkin},
  \citenamefont {Shvets},\ and\ \citenamefont {Fisch}}]{malkin1999fast}%
  \BibitemOpen
  \bibfield  {author} {\bibinfo {author} {\bibfnamefont {V.}~\bibnamefont
  {Malkin}}, \bibinfo {author} {\bibfnamefont {G.}~\bibnamefont {Shvets}}, \
  and\ \bibinfo {author} {\bibfnamefont {N.}~\bibnamefont {Fisch}},\ }\href
  {http://dx.doi.org/10.1103/PhysRevLett.82.4448} {\bibfield  {journal}
  {\bibinfo  {journal} {Physical Review Letters}\ }\textbf {\bibinfo {volume}
  {82}},\ \bibinfo {pages} {4448} (\bibinfo {year} {1999})}\BibitemShut
  {NoStop}%
\bibitem [{\citenamefont {Malkin}\ \emph
  {et~al.}(2000{\natexlab{a}})\citenamefont {Malkin}, \citenamefont {Shvets},\
  and\ \citenamefont {Fisch}}]{malkin2000ultra}%
  \BibitemOpen
  \bibfield  {author} {\bibinfo {author} {\bibfnamefont {V.}~\bibnamefont
  {Malkin}}, \bibinfo {author} {\bibfnamefont {G.}~\bibnamefont {Shvets}}, \
  and\ \bibinfo {author} {\bibfnamefont {N.}~\bibnamefont {Fisch}},\ }\href
  {http://dx.doi.org/10.1063/1.874051} {\bibfield  {journal} {\bibinfo
  {journal} {Physics of Plasmas (1994-present)}\ }\textbf {\bibinfo {volume}
  {7}},\ \bibinfo {pages} {2232} (\bibinfo {year}
  {2000}{\natexlab{a}})}\BibitemShut {NoStop}%
\bibitem [{\citenamefont {Yampolsky}\ \emph {et~al.}(2008)\citenamefont
  {Yampolsky}, \citenamefont {Fisch}, \citenamefont {Malkin}, \citenamefont
  {Valeo}, \citenamefont {Lindberg}, \citenamefont {Wurtele}, \citenamefont
  {Ren}, \citenamefont {Li}, \citenamefont {Morozov},\ and\ \citenamefont
  {Suckewer}}]{yampolsky2008demonstration}%
  \BibitemOpen
  \bibfield  {author} {\bibinfo {author} {\bibfnamefont {N.}~\bibnamefont
  {Yampolsky}}, \bibinfo {author} {\bibfnamefont {N.}~\bibnamefont {Fisch}},
  \bibinfo {author} {\bibfnamefont {V.}~\bibnamefont {Malkin}}, \bibinfo
  {author} {\bibfnamefont {E.}~\bibnamefont {Valeo}}, \bibinfo {author}
  {\bibfnamefont {R.}~\bibnamefont {Lindberg}}, \bibinfo {author}
  {\bibfnamefont {J.}~\bibnamefont {Wurtele}}, \bibinfo {author} {\bibfnamefont
  {J.}~\bibnamefont {Ren}}, \bibinfo {author} {\bibfnamefont {S.}~\bibnamefont
  {Li}}, \bibinfo {author} {\bibfnamefont {A.}~\bibnamefont {Morozov}}, \ and\
  \bibinfo {author} {\bibfnamefont {S.}~\bibnamefont {Suckewer}},\ }\href
  {http://dx.doi.org/10.1063/1.3023153} {\bibfield  {journal} {\bibinfo
  {journal} {Physics of Plasmas (1994-present)}\ }\textbf {\bibinfo {volume}
  {15}},\ \bibinfo {pages} {113104} (\bibinfo {year} {2008})}\BibitemShut
  {NoStop}%
\bibitem [{\citenamefont {Yampolsky}\ and\ \citenamefont
  {Fisch}(2011)}]{yampolsky2011limiting}%
  \BibitemOpen
  \bibfield  {author} {\bibinfo {author} {\bibfnamefont {N.~A.}\ \bibnamefont
  {Yampolsky}}\ and\ \bibinfo {author} {\bibfnamefont {N.~J.}\ \bibnamefont
  {Fisch}},\ }\href {http://dx.doi.org/10.1063/1.3587120} {\bibfield  {journal}
  {\bibinfo  {journal} {Physics of Plasmas (1994-present)}\ }\textbf {\bibinfo
  {volume} {18}},\ \bibinfo {pages} {056711} (\bibinfo {year}
  {2011})}\BibitemShut {NoStop}%
\bibitem [{\citenamefont {Turnbull}\ \emph
  {et~al.}(2012{\natexlab{a}})\citenamefont {Turnbull}, \citenamefont {Li},
  \citenamefont {Morozov},\ and\ \citenamefont
  {Suckewer}}]{turnbull2012simultaneous}%
  \BibitemOpen
  \bibfield  {author} {\bibinfo {author} {\bibfnamefont {D.}~\bibnamefont
  {Turnbull}}, \bibinfo {author} {\bibfnamefont {S.}~\bibnamefont {Li}},
  \bibinfo {author} {\bibfnamefont {A.}~\bibnamefont {Morozov}}, \ and\
  \bibinfo {author} {\bibfnamefont {S.}~\bibnamefont {Suckewer}},\ }\href
  {\doibase 10.1063/1.4748290} {\bibfield  {journal} {\bibinfo  {journal}
  {Physics of Plasmas (1994-present)}\ }\textbf {\bibinfo {volume} {19}},\
  \bibinfo {pages} {083109} (\bibinfo {year} {2012}{\natexlab{a}})}\BibitemShut
  {NoStop}%
\bibitem [{\citenamefont {Turnbull}\ \emph
  {et~al.}(2012{\natexlab{b}})\citenamefont {Turnbull}, \citenamefont {Li},
  \citenamefont {Morozov},\ and\ \citenamefont
  {Suckewer}}]{turnbull2012possible}%
  \BibitemOpen
  \bibfield  {author} {\bibinfo {author} {\bibfnamefont {D.}~\bibnamefont
  {Turnbull}}, \bibinfo {author} {\bibfnamefont {S.}~\bibnamefont {Li}},
  \bibinfo {author} {\bibfnamefont {A.}~\bibnamefont {Morozov}}, \ and\
  \bibinfo {author} {\bibfnamefont {S.}~\bibnamefont {Suckewer}},\ }\href
  {\doibase 10.1063/1.4736856} {\bibfield  {journal} {\bibinfo  {journal}
  {Physics of Plasmas (1994-present)}\ }\textbf {\bibinfo {volume} {19}},\
  \bibinfo {pages} {073103} (\bibinfo {year} {2012}{\natexlab{b}})}\BibitemShut
  {NoStop}%
\bibitem [{\citenamefont {Wu}\ \emph {et~al.}(2016)\citenamefont {Wu},
  \citenamefont {Zhou}, \citenamefont {Zheng}, \citenamefont {Wei},
  \citenamefont {Zhu}, \citenamefont {Su}, \citenamefont {Xie}, \citenamefont
  {Jiao}, \citenamefont {Peng}, \citenamefont {Wang} \emph
  {et~al.}}]{wu2016demonstration}%
  \BibitemOpen
  \bibfield  {author} {\bibinfo {author} {\bibfnamefont {Z.}~\bibnamefont
  {Wu}}, \bibinfo {author} {\bibfnamefont {K.}~\bibnamefont {Zhou}}, \bibinfo
  {author} {\bibfnamefont {X.}~\bibnamefont {Zheng}}, \bibinfo {author}
  {\bibfnamefont {X.}~\bibnamefont {Wei}}, \bibinfo {author} {\bibfnamefont
  {Q.}~\bibnamefont {Zhu}}, \bibinfo {author} {\bibfnamefont {J.}~\bibnamefont
  {Su}}, \bibinfo {author} {\bibfnamefont {N.}~\bibnamefont {Xie}}, \bibinfo
  {author} {\bibfnamefont {Z.}~\bibnamefont {Jiao}}, \bibinfo {author}
  {\bibfnamefont {H.}~\bibnamefont {Peng}}, \bibinfo {author} {\bibfnamefont
  {X.}~\bibnamefont {Wang}},  \emph {et~al.},\ }\href {\doibase
  10.1088/1612-2011/13/10/105301} {\bibfield  {journal} {\bibinfo  {journal}
  {Laser Physics Letters}\ }\textbf {\bibinfo {volume} {13}},\ \bibinfo {pages}
  {105301} (\bibinfo {year} {2016})}\BibitemShut {NoStop}%
\bibitem [{\citenamefont {Ping}\ \emph {et~al.}(2004)\citenamefont {Ping},
  \citenamefont {Cheng}, \citenamefont {Suckewer}, \citenamefont {Clark},\ and\
  \citenamefont {Fisch}}]{ping2004amplification}%
  \BibitemOpen
  \bibfield  {author} {\bibinfo {author} {\bibfnamefont {Y.}~\bibnamefont
  {Ping}}, \bibinfo {author} {\bibfnamefont {W.}~\bibnamefont {Cheng}},
  \bibinfo {author} {\bibfnamefont {S.}~\bibnamefont {Suckewer}}, \bibinfo
  {author} {\bibfnamefont {D.~S.}\ \bibnamefont {Clark}}, \ and\ \bibinfo
  {author} {\bibfnamefont {N.~J.}\ \bibnamefont {Fisch}},\ }\href {\doibase
  10.1103/PhysRevLett.92.175007} {\bibfield  {journal} {\bibinfo  {journal}
  {Physical review letters}\ }\textbf {\bibinfo {volume} {92}},\ \bibinfo
  {pages} {175007} (\bibinfo {year} {2004})}\BibitemShut {NoStop}%
\bibitem [{\citenamefont {Cheng}\ \emph {et~al.}(2005)\citenamefont {Cheng},
  \citenamefont {Avitzour}, \citenamefont {Ping}, \citenamefont {Suckewer},
  \citenamefont {Fisch}, \citenamefont {Hur},\ and\ \citenamefont
  {Wurtele}}]{cheng2005reaching}%
  \BibitemOpen
  \bibfield  {author} {\bibinfo {author} {\bibfnamefont {W.}~\bibnamefont
  {Cheng}}, \bibinfo {author} {\bibfnamefont {Y.}~\bibnamefont {Avitzour}},
  \bibinfo {author} {\bibfnamefont {Y.}~\bibnamefont {Ping}}, \bibinfo {author}
  {\bibfnamefont {S.}~\bibnamefont {Suckewer}}, \bibinfo {author}
  {\bibfnamefont {N.~J.}\ \bibnamefont {Fisch}}, \bibinfo {author}
  {\bibfnamefont {M.~S.}\ \bibnamefont {Hur}}, \ and\ \bibinfo {author}
  {\bibfnamefont {J.~S.}\ \bibnamefont {Wurtele}},\ }\href {\doibase
  10.1103/PhysRevLett.94.045003} {\bibfield  {journal} {\bibinfo  {journal}
  {Physical review letters}\ }\textbf {\bibinfo {volume} {94}},\ \bibinfo
  {pages} {045003} (\bibinfo {year} {2005})}\BibitemShut {NoStop}%
\bibitem [{\citenamefont {Ren}\ \emph {et~al.}(2008)\citenamefont {Ren},
  \citenamefont {Li}, \citenamefont {Morozov}, \citenamefont {Suckewer},
  \citenamefont {Yampolsky}, \citenamefont {Malkin},\ and\ \citenamefont
  {Fisch}}]{ren2008compact}%
  \BibitemOpen
  \bibfield  {author} {\bibinfo {author} {\bibfnamefont {J.}~\bibnamefont
  {Ren}}, \bibinfo {author} {\bibfnamefont {S.}~\bibnamefont {Li}}, \bibinfo
  {author} {\bibfnamefont {A.}~\bibnamefont {Morozov}}, \bibinfo {author}
  {\bibfnamefont {S.}~\bibnamefont {Suckewer}}, \bibinfo {author}
  {\bibfnamefont {N.}~\bibnamefont {Yampolsky}}, \bibinfo {author}
  {\bibfnamefont {V.}~\bibnamefont {Malkin}}, \ and\ \bibinfo {author}
  {\bibfnamefont {N.}~\bibnamefont {Fisch}},\ }\href {\doibase
  10.1063/1.2844352} {\bibfield  {journal} {\bibinfo  {journal} {Physics of
  Plasmas (1994-present)}\ }\textbf {\bibinfo {volume} {15}},\ \bibinfo {pages}
  {056702} (\bibinfo {year} {2008})}\BibitemShut {NoStop}%
\bibitem [{\citenamefont {Palmer}\ \emph {et~al.}(2011)\citenamefont {Palmer},
  \citenamefont {Dover}, \citenamefont {Pogorelsky}, \citenamefont {Babzien},
  \citenamefont {Dudnikova}, \citenamefont {Ispiriyan}, \citenamefont
  {Polyanskiy}, \citenamefont {Schreiber}, \citenamefont {Shkolnikov},
  \citenamefont {Yakimenko},\ and\ \citenamefont
  {Najmudin}}]{palmer2011monoenergetic}%
  \BibitemOpen
  \bibfield  {author} {\bibinfo {author} {\bibfnamefont {C.~A.~J.}\
  \bibnamefont {Palmer}}, \bibinfo {author} {\bibfnamefont {N.~P.}\
  \bibnamefont {Dover}}, \bibinfo {author} {\bibfnamefont {I.}~\bibnamefont
  {Pogorelsky}}, \bibinfo {author} {\bibfnamefont {M.}~\bibnamefont {Babzien}},
  \bibinfo {author} {\bibfnamefont {G.~I.}\ \bibnamefont {Dudnikova}}, \bibinfo
  {author} {\bibfnamefont {M.}~\bibnamefont {Ispiriyan}}, \bibinfo {author}
  {\bibfnamefont {M.~N.}\ \bibnamefont {Polyanskiy}}, \bibinfo {author}
  {\bibfnamefont {J.}~\bibnamefont {Schreiber}}, \bibinfo {author}
  {\bibfnamefont {P.}~\bibnamefont {Shkolnikov}}, \bibinfo {author}
  {\bibfnamefont {V.}~\bibnamefont {Yakimenko}}, \ and\ \bibinfo {author}
  {\bibfnamefont {Z.}~\bibnamefont {Najmudin}},\ }\href {\doibase
  10.1103/PhysRevLett.106.014801} {\bibfield  {journal} {\bibinfo  {journal}
  {Phys. Rev. Lett.}\ }\textbf {\bibinfo {volume} {106}},\ \bibinfo {pages}
  {014801} (\bibinfo {year} {2011})}\BibitemShut {NoStop}%
\bibitem [{\citenamefont {Popmintchev}\ \emph {et~al.}(2012)\citenamefont
  {Popmintchev}, \citenamefont {Chen}, \citenamefont {Popmintchev},
  \citenamefont {Arpin}, \citenamefont {Brown}, \citenamefont
  {Ali{\v{s}}auskas}, \citenamefont {Andriukaitis}, \citenamefont
  {Bal{\v{c}}iunas}, \citenamefont {M{\"u}cke}, \citenamefont {Pugzlys} \emph
  {et~al.}}]{popmintchev2012bright}%
  \BibitemOpen
  \bibfield  {author} {\bibinfo {author} {\bibfnamefont {T.}~\bibnamefont
  {Popmintchev}}, \bibinfo {author} {\bibfnamefont {M.-C.}\ \bibnamefont
  {Chen}}, \bibinfo {author} {\bibfnamefont {D.}~\bibnamefont {Popmintchev}},
  \bibinfo {author} {\bibfnamefont {P.}~\bibnamefont {Arpin}}, \bibinfo
  {author} {\bibfnamefont {S.}~\bibnamefont {Brown}}, \bibinfo {author}
  {\bibfnamefont {S.}~\bibnamefont {Ali{\v{s}}auskas}}, \bibinfo {author}
  {\bibfnamefont {G.}~\bibnamefont {Andriukaitis}}, \bibinfo {author}
  {\bibfnamefont {T.}~\bibnamefont {Bal{\v{c}}iunas}}, \bibinfo {author}
  {\bibfnamefont {O.~D.}\ \bibnamefont {M{\"u}cke}}, \bibinfo {author}
  {\bibfnamefont {A.}~\bibnamefont {Pugzlys}},  \emph {et~al.},\ }\href
  {\doibase 10.1126/science.1218497} {\bibfield  {journal} {\bibinfo  {journal}
  {Science}\ }\textbf {\bibinfo {volume} {336}},\ \bibinfo {pages} {1287}
  (\bibinfo {year} {2012})}\BibitemShut {NoStop}%
\bibitem [{\citenamefont {Pigeon}(2014)}]{pigeon2014generation}%
  \BibitemOpen
  \bibfield  {author} {\bibinfo {author} {\bibfnamefont {J.}~\bibnamefont
  {Pigeon}},\ }\emph {\bibinfo {title} {Generation of ultra-broadband, mid-{IR}
  radiation in {GaAs} pumped by picosecond 10 $\mu$m laser pulses}},\ \href
  {https://escholarship.org/uc/item/0zh483z5} {Master's thesis},\ \bibinfo
  {school} {UCLA: Electrical Engineering} (\bibinfo {year} {2014})\BibitemShut
  {NoStop}%
\bibitem [{\citenamefont {Pigeon}\ \emph {et~al.}(2015)\citenamefont {Pigeon},
  \citenamefont {Tochitsky},\ and\ \citenamefont {Joshi}}]{pigeon2015high}%
  \BibitemOpen
  \bibfield  {author} {\bibinfo {author} {\bibfnamefont {J.}~\bibnamefont
  {Pigeon}}, \bibinfo {author} {\bibfnamefont {S.~Y.}\ \bibnamefont
  {Tochitsky}}, \ and\ \bibinfo {author} {\bibfnamefont {C.}~\bibnamefont
  {Joshi}},\ }\href {\doibase 10.1364/OL.40.005730} {\bibfield  {journal}
  {\bibinfo  {journal} {Optics letters}\ }\textbf {\bibinfo {volume} {40}},\
  \bibinfo {pages} {5730} (\bibinfo {year} {2015})}\BibitemShut {NoStop}%
\bibitem [{\citenamefont {Mitrofanov}\ \emph {et~al.}(2015)\citenamefont
  {Mitrofanov}, \citenamefont {Voronin}, \citenamefont {Sidorov-Biryukov},
  \citenamefont {Pug{\v{z}}lys}, \citenamefont {Stepanov}, \citenamefont
  {Andriukaitis}, \citenamefont {Fl{\"o}ry}, \citenamefont {Ali{\v{s}}auskas},
  \citenamefont {Fedotov}, \citenamefont {Baltu{\v{s}}ka} \emph
  {et~al.}}]{mitrofanov2015mid}%
  \BibitemOpen
  \bibfield  {author} {\bibinfo {author} {\bibfnamefont {A.}~\bibnamefont
  {Mitrofanov}}, \bibinfo {author} {\bibfnamefont {A.}~\bibnamefont {Voronin}},
  \bibinfo {author} {\bibfnamefont {D.}~\bibnamefont {Sidorov-Biryukov}},
  \bibinfo {author} {\bibfnamefont {A.}~\bibnamefont {Pug{\v{z}}lys}}, \bibinfo
  {author} {\bibfnamefont {E.}~\bibnamefont {Stepanov}}, \bibinfo {author}
  {\bibfnamefont {G.}~\bibnamefont {Andriukaitis}}, \bibinfo {author}
  {\bibfnamefont {T.}~\bibnamefont {Fl{\"o}ry}}, \bibinfo {author}
  {\bibfnamefont {S.}~\bibnamefont {Ali{\v{s}}auskas}}, \bibinfo {author}
  {\bibfnamefont {A.}~\bibnamefont {Fedotov}}, \bibinfo {author} {\bibfnamefont
  {A.}~\bibnamefont {Baltu{\v{s}}ka}},  \emph {et~al.},\ }\href {\doibase
  10.1038/srep08368} {\bibfield  {journal} {\bibinfo  {journal} {Scientific
  Reports}\ }\textbf {\bibinfo {volume} {5}} (\bibinfo {year} {2015}),\
  10.1038/srep08368}\BibitemShut {NoStop}%
\bibitem [{\citenamefont {Voronin}\ \emph {et~al.}(2016)\citenamefont
  {Voronin}, \citenamefont {Lanin},\ and\ \citenamefont
  {Zheltikov}}]{voronin2016modeling}%
  \BibitemOpen
  \bibfield  {author} {\bibinfo {author} {\bibfnamefont {A.}~\bibnamefont
  {Voronin}}, \bibinfo {author} {\bibfnamefont {A.}~\bibnamefont {Lanin}}, \
  and\ \bibinfo {author} {\bibfnamefont {A.}~\bibnamefont {Zheltikov}},\ }\href
  {\doibase 10.1364/OE.24.023207} {\bibfield  {journal} {\bibinfo  {journal}
  {Optics Express}\ }\textbf {\bibinfo {volume} {24}},\ \bibinfo {pages}
  {23207} (\bibinfo {year} {2016})}\BibitemShut {NoStop}%
\bibitem [{\citenamefont {Cerullo}\ and\ \citenamefont
  {De~Silvestri}(2003)}]{cerullo2003ultrafast}%
  \BibitemOpen
  \bibfield  {author} {\bibinfo {author} {\bibfnamefont {G.}~\bibnamefont
  {Cerullo}}\ and\ \bibinfo {author} {\bibfnamefont {S.}~\bibnamefont
  {De~Silvestri}},\ }\href {\doibase 10.1063/1.1523642} {\bibfield  {journal}
  {\bibinfo  {journal} {Review of scientific instruments}\ }\textbf {\bibinfo
  {volume} {74}},\ \bibinfo {pages} {1} (\bibinfo {year} {2003})}\BibitemShut
  {NoStop}%
\bibitem [{\citenamefont {Sell}\ \emph {et~al.}(2008)\citenamefont {Sell},
  \citenamefont {Leitenstorfer},\ and\ \citenamefont {Huber}}]{sell2008phase}%
  \BibitemOpen
  \bibfield  {author} {\bibinfo {author} {\bibfnamefont {A.}~\bibnamefont
  {Sell}}, \bibinfo {author} {\bibfnamefont {A.}~\bibnamefont {Leitenstorfer}},
  \ and\ \bibinfo {author} {\bibfnamefont {R.}~\bibnamefont {Huber}},\ }\href
  {\doibase 10.1364/OL.33.002767} {\bibfield  {journal} {\bibinfo  {journal}
  {Optics Letters}\ }\textbf {\bibinfo {volume} {33}},\ \bibinfo {pages} {2767}
  (\bibinfo {year} {2008})}\BibitemShut {NoStop}%
\bibitem [{\citenamefont {Polyanskiy}\ \emph {et~al.}(2011)\citenamefont
  {Polyanskiy}, \citenamefont {Pogorelsky},\ and\ \citenamefont
  {Yakimenko}}]{polyanskiy2011picosecond}%
  \BibitemOpen
  \bibfield  {author} {\bibinfo {author} {\bibfnamefont {M.~N.}\ \bibnamefont
  {Polyanskiy}}, \bibinfo {author} {\bibfnamefont {I.~V.}\ \bibnamefont
  {Pogorelsky}}, \ and\ \bibinfo {author} {\bibfnamefont {V.}~\bibnamefont
  {Yakimenko}},\ }\href {\doibase 10.1364/OE.19.007717} {\bibfield  {journal}
  {\bibinfo  {journal} {Optics Express}\ }\textbf {\bibinfo {volume} {19}},\
  \bibinfo {pages} {7717} (\bibinfo {year} {2011})}\BibitemShut {NoStop}%
\bibitem [{\citenamefont {Pogorelsky}\ \emph {et~al.}(2015)\citenamefont
  {Pogorelsky}, \citenamefont {Babzien}, \citenamefont {Ben-Zvi}, \citenamefont
  {Skaritka},\ and\ \citenamefont {Polyanskiy}}]{pogorelsky2015bestia}%
  \BibitemOpen
  \bibfield  {author} {\bibinfo {author} {\bibfnamefont {I.~V.}\ \bibnamefont
  {Pogorelsky}}, \bibinfo {author} {\bibfnamefont {M.}~\bibnamefont {Babzien}},
  \bibinfo {author} {\bibfnamefont {I.}~\bibnamefont {Ben-Zvi}}, \bibinfo
  {author} {\bibfnamefont {J.}~\bibnamefont {Skaritka}}, \ and\ \bibinfo
  {author} {\bibfnamefont {M.~N.}\ \bibnamefont {Polyanskiy}},\ }\href
  {\doibase 10.1016/j.nima.2015.11.126} {\bibfield  {journal} {\bibinfo
  {journal} {Nuclear Instruments and Methods in Physics Research Section A:
  Accelerators, Spectrometers, Detectors and Associated Equipment}\ } (\bibinfo
  {year} {2015}),\ 10.1016/j.nima.2015.11.126}\BibitemShut {NoStop}%
\bibitem [{\citenamefont {Kruer}(1988)}]{kruer1988physics}%
  \BibitemOpen
  \bibfield  {author} {\bibinfo {author} {\bibfnamefont {W.~L.}\ \bibnamefont
  {Kruer}},\ }\href@noop {} {\emph {\bibinfo {title} {The physics of laser
  plasma interactions}}}\ (\bibinfo  {publisher} {Reading, MA (US);
  Addison-Wesley Publishing Co.},\ \bibinfo {year} {1988})\BibitemShut
  {NoStop}%
\bibitem [{\citenamefont {Toroker}\ \emph {et~al.}(2014)\citenamefont
  {Toroker}, \citenamefont {Malkin},\ and\ \citenamefont
  {Fisch}}]{toroker2014backward}%
  \BibitemOpen
  \bibfield  {author} {\bibinfo {author} {\bibfnamefont {Z.}~\bibnamefont
  {Toroker}}, \bibinfo {author} {\bibfnamefont {V.}~\bibnamefont {Malkin}}, \
  and\ \bibinfo {author} {\bibfnamefont {N.}~\bibnamefont {Fisch}},\ }\href
  {\doibase 10.1063/1.4902362} {\bibfield  {journal} {\bibinfo  {journal}
  {Physics of Plasmas (1994-present)}\ }\textbf {\bibinfo {volume} {21}},\
  \bibinfo {pages} {113110} (\bibinfo {year} {2014})}\BibitemShut {NoStop}%
\bibitem [{\citenamefont {Edwards}\ \emph {et~al.}(2015)\citenamefont
  {Edwards}, \citenamefont {Toroker}, \citenamefont {Mikhailova},\ and\
  \citenamefont {Fisch}}]{edwards2015efficiency}%
  \BibitemOpen
  \bibfield  {author} {\bibinfo {author} {\bibfnamefont {M.~R.}\ \bibnamefont
  {Edwards}}, \bibinfo {author} {\bibfnamefont {Z.}~\bibnamefont {Toroker}},
  \bibinfo {author} {\bibfnamefont {J.~M.}\ \bibnamefont {Mikhailova}}, \ and\
  \bibinfo {author} {\bibfnamefont {N.~J.}\ \bibnamefont {Fisch}},\ }\href
  {\doibase 10.1063/1.4926514} {\bibfield  {journal} {\bibinfo  {journal}
  {Physics of Plasmas (1994-present)}\ }\textbf {\bibinfo {volume} {22}},\
  \bibinfo {pages} {074501} (\bibinfo {year} {2015})}\BibitemShut {NoStop}%
\bibitem [{\citenamefont {Ren}\ \emph {et~al.}(2007)\citenamefont {Ren},
  \citenamefont {Cheng}, \citenamefont {Li},\ and\ \citenamefont
  {Suckewer}}]{ren2007new}%
  \BibitemOpen
  \bibfield  {author} {\bibinfo {author} {\bibfnamefont {J.}~\bibnamefont
  {Ren}}, \bibinfo {author} {\bibfnamefont {W.}~\bibnamefont {Cheng}}, \bibinfo
  {author} {\bibfnamefont {S.}~\bibnamefont {Li}}, \ and\ \bibinfo {author}
  {\bibfnamefont {S.}~\bibnamefont {Suckewer}},\ }\href {\doibase
  10.1038/nphys717} {\bibfield  {journal} {\bibinfo  {journal} {Nature
  Physics}\ }\textbf {\bibinfo {volume} {3}},\ \bibinfo {pages} {732} (\bibinfo
  {year} {2007})}\BibitemShut {NoStop}%
\bibitem [{\citenamefont {Strozzi}(2005)}]{strozzi2005vlasov}%
  \BibitemOpen
  \bibfield  {author} {\bibinfo {author} {\bibfnamefont {D.~J.}\ \bibnamefont
  {Strozzi}},\ }\emph {\bibinfo {title} {Vlasov simulations of kinetic
  enhancement of Raman backscatter in laser fusion plasmas}},\ \href
  {http://hdl.handle.net/1721.1/34974} {Ph.D. thesis},\ \bibinfo  {school}
  {Massachusetts Institute of Technology} (\bibinfo {year} {2005})\BibitemShut
  {NoStop}%
\bibitem [{\citenamefont {Malkin}\ and\ \citenamefont
  {Fisch}(2014)}]{malkin2014key}%
  \BibitemOpen
  \bibfield  {author} {\bibinfo {author} {\bibfnamefont {V.}~\bibnamefont
  {Malkin}}\ and\ \bibinfo {author} {\bibfnamefont {N.}~\bibnamefont {Fisch}},\
  }\href {\doibase 10.1140/epjst/e2014-02168-0} {\bibfield  {journal} {\bibinfo
   {journal} {The European Physical Journal Special Topics}\ }\textbf {\bibinfo
  {volume} {223}},\ \bibinfo {pages} {1157} (\bibinfo {year}
  {2014})}\BibitemShut {NoStop}%
\bibitem [{\citenamefont {Gordon}\ \emph {et~al.}(2000)\citenamefont {Gordon},
  \citenamefont {Mori},\ and\ \citenamefont {Antonsen}}]{gordon2000}%
  \BibitemOpen
  \bibfield  {author} {\bibinfo {author} {\bibfnamefont {D.~F.}\ \bibnamefont
  {Gordon}}, \bibinfo {author} {\bibfnamefont {W.}~\bibnamefont {Mori}}, \ and\
  \bibinfo {author} {\bibfnamefont {T.~M.}\ \bibnamefont {Antonsen}},\ }\href
  {\doibase 10.1109/27.893300} {\bibfield  {journal} {\bibinfo  {journal} {IEEE
  Transactions on Plasma Science}\ }\textbf {\bibinfo {volume} {28}},\ \bibinfo
  {pages} {1135} (\bibinfo {year} {2000})}\BibitemShut {NoStop}%
\bibitem [{\citenamefont {Malkin}\ \emph
  {et~al.}(2000{\natexlab{b}})\citenamefont {Malkin}, \citenamefont {Shvets},\
  and\ \citenamefont {Fisch}}]{malkin2000detuned}%
  \BibitemOpen
  \bibfield  {author} {\bibinfo {author} {\bibfnamefont {V.}~\bibnamefont
  {Malkin}}, \bibinfo {author} {\bibfnamefont {G.}~\bibnamefont {Shvets}}, \
  and\ \bibinfo {author} {\bibfnamefont {N.}~\bibnamefont {Fisch}},\ }\href
  {\doibase 10.1103/PhysRevLett.84.1208} {\bibfield  {journal} {\bibinfo
  {journal} {Physical Review Letters}\ }\textbf {\bibinfo {volume} {84}},\
  \bibinfo {pages} {1208} (\bibinfo {year} {2000}{\natexlab{b}})}\BibitemShut
  {NoStop}%
\bibitem [{\citenamefont {Catto}\ and\ \citenamefont
  {Speziale}(1977)}]{catto1977strong}%
  \BibitemOpen
  \bibfield  {author} {\bibinfo {author} {\bibfnamefont {P.}~\bibnamefont
  {Catto}}\ and\ \bibinfo {author} {\bibfnamefont {T.}~\bibnamefont
  {Speziale}},\ }\href {\doibase 10.1063/1.861688} {\bibfield  {journal}
  {\bibinfo  {journal} {Physics of Fluids (1958-1988)}\ }\textbf {\bibinfo
  {volume} {20}},\ \bibinfo {pages} {167} (\bibinfo {year} {1977})}\BibitemShut
  {NoStop}%
\bibitem [{\citenamefont {Huba}(2004)}]{huba2004nrl}%
  \BibitemOpen
  \bibfield  {author} {\bibinfo {author} {\bibfnamefont {J.~D.}\ \bibnamefont
  {Huba}},\ }\href@noop {} {\emph {\bibinfo {title} {{NRL}: {P}lasma
  formulary}}},\ \bibinfo {type} {Tech. Rep.}\ (\bibinfo  {institution} {DTIC
  Document},\ \bibinfo {year} {2004})\BibitemShut {NoStop}%
\bibitem [{\citenamefont {Fitzpatrick}(2014)}]{fitzpatrick2014plasma}%
  \BibitemOpen
  \bibfield  {author} {\bibinfo {author} {\bibfnamefont {R.}~\bibnamefont
  {Fitzpatrick}},\ }\href@noop {} {\emph {\bibinfo {title} {Plasma physics: an
  introduction}}}\ (\bibinfo  {publisher} {CRC Press},\ \bibinfo {year}
  {2014})\BibitemShut {NoStop}%
\bibitem [{\citenamefont {Waskom}\ \emph {et~al.}(2016)\citenamefont {Waskom},
  \citenamefont {Botvinnik}, \citenamefont {drewokane}, \citenamefont {Hobson},
  \citenamefont {Halchenko}, \citenamefont {Lukauskas}, \citenamefont
  {Warmenhoven}, \citenamefont {Cole}, \citenamefont {Hoyer}, \citenamefont
  {Vanderplas},\ and\ \citenamefont {et~al.}}]{seaborn}%
  \BibitemOpen
  \bibfield  {author} {\bibinfo {author} {\bibfnamefont {M.}~\bibnamefont
  {Waskom}}, \bibinfo {author} {\bibfnamefont {O.}~\bibnamefont {Botvinnik}},
  \bibinfo {author} {\bibnamefont {drewokane}}, \bibinfo {author}
  {\bibfnamefont {P.}~\bibnamefont {Hobson}}, \bibinfo {author} {\bibfnamefont
  {Y.}~\bibnamefont {Halchenko}}, \bibinfo {author} {\bibfnamefont
  {S.}~\bibnamefont {Lukauskas}}, \bibinfo {author} {\bibfnamefont
  {J.}~\bibnamefont {Warmenhoven}}, \bibinfo {author} {\bibfnamefont {J.~B.}\
  \bibnamefont {Cole}}, \bibinfo {author} {\bibfnamefont {S.}~\bibnamefont
  {Hoyer}}, \bibinfo {author} {\bibfnamefont {J.}~\bibnamefont {Vanderplas}}, \
  and\ \bibinfo {author} {\bibnamefont {et~al.}},\ }\href {\doibase
  10.5281/zenodo.45133} {\enquote {\bibinfo {title} {seaborn: v0.7.0 (january
  2016)},}\ } (\bibinfo {year} {2016})\BibitemShut {NoStop}%
\bibitem [{\citenamefont {McKinney}(2010)}]{pandas}%
  \BibitemOpen
  \bibfield  {author} {\bibinfo {author} {\bibfnamefont {W.}~\bibnamefont
  {McKinney}},\ }in\ \href@noop {} {\emph {\bibinfo {booktitle} {Proceedings of
  the 9th Python in Science Conference}}},\ \bibinfo {editor} {edited by\
  \bibinfo {editor} {\bibfnamefont {S.}~\bibnamefont {van~der Walt}}\ and\
  \bibinfo {editor} {\bibfnamefont {J.}~\bibnamefont {Millman}}}\ (\bibinfo
  {year} {2010})\ pp.\ \bibinfo {pages} {51 -- 56}\BibitemShut {NoStop}%
\bibitem [{\citenamefont {Van Der~Walt}\ \emph {et~al.}(2011)\citenamefont {Van
  Der~Walt}, \citenamefont {Colbert},\ and\ \citenamefont {Varoquaux}}]{numpy}%
  \BibitemOpen
  \bibfield  {author} {\bibinfo {author} {\bibfnamefont {S.}~\bibnamefont {Van
  Der~Walt}}, \bibinfo {author} {\bibfnamefont {S.~C.}\ \bibnamefont
  {Colbert}}, \ and\ \bibinfo {author} {\bibfnamefont {G.}~\bibnamefont
  {Varoquaux}},\ }\href@noop {} {\bibfield  {journal} {\bibinfo  {journal}
  {Computing in Science \& Engineering}\ }\textbf {\bibinfo {volume} {13}},\
  \bibinfo {pages} {22} (\bibinfo {year} {2011})}\BibitemShut {NoStop}%
\bibitem [{\citenamefont {Johansson}\ \emph {et~al.}(2013)\citenamefont
  {Johansson} \emph {et~al.}}]{mpmath}%
  \BibitemOpen
  \bibfield  {author} {\bibinfo {author} {\bibfnamefont {F.}~\bibnamefont
  {Johansson}} \emph {et~al.},\ }\href@noop {} {\emph {\bibinfo {title}
  {mpmath: a {P}ython library for arbitrary-precision floating-point arithmetic
  (version 0.18)}}} (\bibinfo {year} {2013}),\ \bibinfo {note} {{\tt
  http://mpmath.org/}}\BibitemShut {NoStop}%
\end{thebibliography}%

\iftoggle{showgit}{
    \pagestyle{lastpage}
}{
    \pagestyle{body}
}

\end{document}